\documentclass[letterpaper,twocolumn,10pt]{article}
\usepackage{usenix2019_v3}

\usepackage{tikz}
\usepackage{amsmath}
\usepackage[available,functional,reproduced]{usenixbadges/usenixbadges}
\usepackage{gauntlet_macros}

\begin{document}
%-------------------------------------------------------------------------------

%don't want date printed
\date{}

\title{\Large \bf \toolname: Finding Bugs in Compilers for Programmable Packet Processing}

\author{
{\rm Fabian Ruffy, Tao Wang, and Anirudh Sivaraman}\\
New York University
} % end author
\maketitle

\begin{sloppypar}
\begin{abstract}
Programmable packet-processing devices such as programmable switches and network interface cards are becoming mainstream. These devices are configured in a domain-specific language such as P4, using a compiler to translate packet-processing programs into instructions for different targets. As networks with programmable devices become widespread, it is critical that these compilers be dependable.

This paper considers the problem of finding bugs in compilers for packet processing in the context of P4\textsubscript{16}. We introduce domain-specific techniques to induce both abnormal termination of the compiler (crash bugs) and miscompilation (semantic bugs). We apply these techniques to (1) the open-source P4 compiler (\pfourc) infrastructure, which serves as a common base for different P4 back ends; (2) the P4 back end for the P4 reference software switch; and (3) the P4 back end for the Barefoot Tofino switch.

Across the 3 platforms, over 8 months of bug finding, our tool \toolname detected \bugtotal new  and distinct bugs (\bugcrash crash and \bugsemantic semantic), which we confirmed with the respective compiler developers. \bugfixed have been fixed (\bugfixedcrash crash and \bugfixedsemantic semantic); the remaining have been assigned to a developer. Our bug-finding efforts also led to \specchanges P4 specification changes. We have open sourced \toolname at \projecturl and it now runs within \pfourc's continuous integration pipeline.
\end{abstract}
\section{Introduction}
\label{sec:introduction}
Programmable packet-processing devices in the form of programmable switches and network interface cards (NICs) are now common. Such devices provide network flexibility, allowing network operators to customize their network, researchers to experiment with new network algorithms, and equipment vendors to upgrade features rapidly in firmware rather than waiting for new hardware. At the core of this move to programmable packet processing are the domain-specific languages (DSLs) for packet processing, along with the compilers that compile DSL programs.

Several commercial products now use such DSLs for packet processing. For instance, Intel~\cite{tofino}, Broadcom~\cite{trident}, Nvidia~\cite{nvidia_doca_sdk}, and Cisco~\cite{silicon_one} have switches and NICs programmable in DSLs such as NPL~\cite{npl} and P4~\cite{p4}. Other efforts (e.g., from Google and the Open Networking Foundation (ONF)) use the P4 language to model the behavior of fixed-function devices~\cite{p4_google_stratum}.

These devices, whether fixed or programmable, are a critical part of the network infrastructure because they process every packet going through the network. Hence, a miscompiled program can persistently affect packet processing. It can also be very hard to track down miscompilations due to the lack of sophisticated debugging support on these devices. As network programmability becomes increasingly common, these DSL compilers will need to be as dependable as general-purpose compilers such as GCC and LLVM.

Motivated by these concerns, this paper considers the problem of finding bugs in compilers for packet processing. Because of the large open-source community around it, we build our work on the P4~\cite{p4} language, but our ideas also extend to similar DSLs such as NPL~\cite{npl}.

Bug finding in compilers is a well-studied topic, especially in the context of C~\cite{csmith, emi, equality_saturation, dol, athena}. Past approaches (\S\ref{sec:background}) to bug finding in C compilers include fuzz testing by using randomly generated C programs~\cite{csmith, emi}, translation validation (i.e., proving that a compiler correctly translated a given input program to an output program)~\cite{tv_optimizing, tv_original}, and verification of individual compiler passes~\cite{alive}. These prior approaches have to contend with many difficulties inherent to a general-purpose language like C, e.g., generating random programs that avoid undefined and unspecified behavior~\cite{csmith, emi},  providing semantics for pointers and memory aliasing~\cite{alive}, and inferring loop invariants and simulation relations to successfully perform translation validation~\cite{tv_original}.

Our key insight is that the restricted nature of a DSL such as P4 allows us to avoid much of the complexity associated with bug finding in general-purpose language compilers. In particular, the simpler nature of P4 (e.g., no loops or pointers) allowed us to more easily develop formal semantics, which can then be used as the basis for both automated high-accuracy translation validation and model-based testing~\cite{model_based_testing}. We leverage this insight to build a compiler bug-finding tool for P4 called \toolname. \toolname uses three key ideas: random program generation, translation validation, and model-based testing. We now describe these ideas and show how the restrictions of P4 allows them to be simpler than prior work.

First, we use random program generation (\S\ref{sec:rand_prog}) to produce syntactically correct and well-typed P4 programs that still induce P4 compiler crashes. Because P4 has very little undefined behavior~\cite[\S 7.1.6]{p416_spec}, random program generation is considerably simpler for P4 than for C~\cite{csmith}. The generator does not have to painstakingly avoid generating programs with undefined and unspecified behavior, which can be interpreted differently across different compilers. The smaller and simpler grammar of P4 relative to C also simplifies the development of a random program generator.

Second, we use translation validation (\S\ref{sec:tv})~\cite{tv_original, tv_optimizing} to find miscompilations in P4 compilers in which we can access the transformed program after every compiler pass. Translation validation has been used in the context of C compilers before, but has suffered one of two shortcomings. It either needs considerable manual effort per compiler pass (e.g., Crellvm~\cite{crellvm} requires several 100 lines of manual proof-generation code for each pass; Alive~\cite{alive} requires manual translation of optimizations into the Alive DSL) or suffers from a small rate of false positives and false negatives (e.g.,~\cite{symdiff, tv_optimizing}). Fundamentally, this is inevitable for unrestricted C: proving program equivalence in the presence of unbounded loops is undecidable. In our case, however, the finite nature of P4\footnote{Finite in that input and output packets and state are finite bit vectors. Loops are bounded (parsing~\cite[\S 12]{p416_spec}) or forbidden (control flow~\cite[\S 13]{p416_spec}).} makes P4 program equivalence decidable and addresses both shortcomings. Thus, our use of translation validation is both precise and fully automated, requiring manual effort only to develop semantics for the P4 language---not manual effort per compiler pass.

Third, we use model-based testing (\S\ref{sec:symbex}) to generate input-output test packets for P4 programs based on the semantics we had to develop for translation validation. We use these test packet pairs to find miscompilations in black-box and proprietary P4 compilers where we can not access the transformed program after every compiler pass. Testing for general-purpose languages~\cite{klee} is effective at generating inputs that provide sufficient path coverage by finding inputs satisfying  path conditions. But without language semantics, determining the correct output for these test inputs is hard. By creating formal semantics for P4 for translation validation, we are able to generate both input and output test packets, which can then be used to test the implementation produced by the compiler for a P4 program.

We applied \toolname to 3 platforms (\S\ref{sec:results}): (1) the open-source P4 compiler infrastructure (\pfourc)~\cite{p416}, which serves as a common base for different P4 compiler implementations; (2) the P4 back end for the open-source P4 behavioral model (\bmv)~\cite{bmv2}, a reference software switch for P4; and (3) the P4 back end for Barefoot Tofino, a high-speed programmable switching chip~\cite{tofino}. Across these 3 platforms, and over 8 months of testing, we found a total of \bugtotal new and distinct bugs, all of which were confirmed and assigned to a compiler developer. Our efforts also led to \specchanges changes~\cite[\S A.1]{p416_spec} to the P4 specification. \bugfixed of these bugs have already been fixed. We analyze these bugs in detail and describe where they were found, their root causes, and which commits introduced them. \toolname has been merged into the continuous integration pipeline of the official P4 reference compiler~\cite{gauntlet_ci}. Our tools are open source and available at \projecturl. To our knowledge, Gauntlet is the first example of using translation validation for compiler bug finding on a production compiler as part of its continuous integration workflow.

While Gauntlet has been very effective, it is still restricted in the kinds of bugs, compiler passes, and language constructs it can handle. We describe these restrictions to motivate future work (\S\ref{sec:future_work}). Further, while we developed these bug-finding techniques in the context of P4, we believe the lessons we have learned (\S\ref{ss:lessons}) apply beyond P4 to other DSLs with simpler semantics relative to general-purpose languages (e.g., the HLO IR for the TensorFlow~\cite{tensorflow} XLA compiler~\cite{xla}).
\section{Background and Motivation}
\label{sec:background}
\subsection{Approaches to Testing Compilers}
\Para{Levels of compiler testing.} A compiler must reject incorrect programs with an appropriate error message and accurately translate correct programs. However, a program can be correct to varying levels. McKeeman~\cite{differential_testing} provides a taxonomy of these levels in the context of C (Table~\ref{tab:diff_test}). Each level corresponds to the program getting deeper into the compiler before it is rejected (e.g., lexer, parser, type checker, optimizer, code generator). The difficulty of generating test programs also goes up with increasing input level. For instance, while general-purpose fuzzers such as AFL~\cite{afl} are sufficient to stress test the lexer, more sophistication is required to generate syntactically correct and well-typed programs, which are required to test the optimizer. In the context of the P4 compiler, we observed very limited success in bug finding using a general-purpose fuzzer such as AFL. This is because testing at the first few levels of Table~\ref{tab:diff_test} is already handled adequately by P4's open-source compiler test suite~\cite[\S 3.4]{p416}.

Hence, for this paper, we only consider programs at the higher levels: static, dynamic, and model-conforming. These are programs that pass the lexing, parsing, type checking, and semantic analysis phases of the compiler, but still trigger compiler bugs. Like Csmith~\cite{csmith}, we categorize bugs into {\em crash bugs} and {\em semantic bugs}. A crash bug occurs when the compiler abnormally terminates on an input program without producing either an output program or a useful error message. Crash bugs include segmentation faults, assertion violations, incomplete error messages, and out-of-memory errors. A semantic bug occurs when the compiler produces an output executable, but the executable's behavior is different from the input program, e.g., due to an incorrect program transformation in a compiler optimization pass. In P4, semantic bugs manifest as any packet output that differs from the expected packet output given an input packet. Crash bugs we are interested in correspond to level 5 in Table~\ref{tab:diff_test}; semantic bugs correspond to levels 6 and 7.

%%%%%%%%%%%%%%%%%%% TABLE BEGIN
\begin{table}[!t]
\small
      \centering
      \resizebox{\columnwidth}{!}{
      \begin{tabular}{lll}
      \hline
       Level & Input Class & Example of incorrect input \\
      \hline
      \hline
      1 & Sequence  of ASCII characters & Binary files \\
      \hline
      2 & Sequence of words and spaces & Variable name beginning with \$ \\
      \hline
      3 & Syntactically correct & Missing semicolon \\
      \hline
      4 & Type correct & Adding int to string \\
      \hline
      5 & Statically conforming & Undefined variables \\
      \hline
      6 & Dynamically conforming & Program throwing exceptions \\
      \hline
      7 & Model-conforming & Program producing wrong outputs \\
      \hline
      \end{tabular}
      }
      \caption{McKeeman's~\cite{differential_testing} 7 levels of C compiler correctness.}
      \label{tab:diff_test}
\end{table}

%%%%%%%%%%%%%%%%%%% TABLE BEGIN

\Para{Bug-finding strategies.} We now look at how compiler bugs are found. A key challenge in compiler bug finding is the oracle problem. Given an input program to a compiler, the expected outcome (i.e., should it accept/reject the program and what should the output be?) is unclear unless one consults an all-knowing oracle. Below, we outline the major techniques used to approximate this oracle knowledge.

In differential testing~\cite{differential_testing}, given two compilers, which both receive the same input program, if compiler A's output (after compiling and running the program) differs from compiler B's output, there is a bug in one of them. This works as long as there are at least two independent compiler implementations for the same language. Csmith~\cite{csmith} is one example of this approach; it feeds the same randomly generated C program to multiple C compilers and checks whether the outputs generated by executing the binary produced by each compiler differ. Another example is Different Optimization Levels (DOL)~\cite{dol}, which selectively omits compiler optimizations and compares compiler outputs with and without these optimization passes. If the end result differs after specific passes have been skipped or added, it points to a bug. This technique can be used in any compiler framework that supports selective omission of optimizations.

Metamorphic testing~\cite{metamorphic_testing} can serve a similar role as differential testing, especially when multiple compilers are not readily available or optimization passes can not be easily disabled. Instead of feeding the same input program to different compilers, different input programs that are expected to produce the same compiler output are fed to the same compiler. The run-time outputs after compiling these different input programs are compared to determine if there is a bug or not. EMI is an example of this approach~\cite{emi}. Given a randomly generated C program $P$, and random input $I$ to this program, EMI uses the path coverage tool \texttt{gcov}~\cite{gcov} to identify dead code in $P$ when run on input $I$. EMI then prunes away this dead code to produce new programs $P'$ whose output must agree with $P$'s output when run on the input $I$. Then EMI compiles and runs both $P$ and $P'$ to check whether they indeed produce the same output when given $I$ as input.

Translation validation is a bug-finding technique that converts the program before and after a compiler optimization pass into a logical formula and checks if both programs/formulas are equivalent using a constraint solver~\cite{tv_optimizing, tv_original, alive, voc}. A failed check indicates a semantic bug. Program equivalence is an undecidable problem for Turing-complete languages such as C, requiring manual assistance to perform translation validation. Typical examples of manual assistance are (1) simulation relations, which encode correspondences between variables in two programs; and (2) loop invariants, required to prove the equivalence of programs with loops. While it is possible to just unroll loops a constant number of times~\cite{symdiff} or learn these relations~\cite{ddec, tv_optimizing}, these techniques are not guaranteed to be precise and occasionally generate false alarms~\cite{crellvm}. The occurrence of false alarms makes translation validation an unlikely choice for recurring use in compiler testing for general-purpose languages (e.g., for continuous integration). This is  because the number of false alarms typically exceeds compiler developer tolerance.

\subsection{Motivating \toolname's Design}
\Para{Random program generation for crash bugs.} From EMI and Csmith, we borrow the idea of generating random programs that are lexically, syntactically, and semantically correct. Unlike EMI and Csmith, however, our random program generation is simpler. It does not have to avoid undefined behavior, which, by design, is quite limited in P4\textsubscript{16}. Further, generating programs with undefined behavior helps us flag compiler passes that might exploit undefined behavior in counter-intuitive ways~\cite{stack}. We feed these randomly generated programs to the compiler to see if it generates a crash, typically a failure of an assertion written by the P4 compiler developers.

\Para{Translation validation for semantic bugs.} Differential and metamorphic testing allow us to compare different run-time outputs from compiled programs to detect semantic bugs. However, we can not directly apply either to P4 compilers. Differential testing requires two or more independent compiler implementations that are comparable in their output. P4\textsubscript{16} compilers for different hardware and software targets are not comparable because program behavior is target-dependent~\cite[\S 2.1]{p416}. Presently there aren't multiple independent compilers for the same target. Developing an entirely new compiler exclusively for the sake of testing the existing compiler is not productive because it can only be reused for one target. Metamorphic testing~\cite{emi}, on the other hand, requires the use of code-coverage tools such as \texttt{gcov} to determine which parts of the program are touched by a given input. Concurrent research~\cite{ball_larus_p4} has proposed such tools for P4, but these tools were not available when we commenced work on \toolname.

On the other hand, P4's domain-specific restrictions make translation validation easier relative to general-purpose languages such as C. P4 programs are finite-state and finite-time, which makes program equivalence decidable at a theoretical level. At the practical level, P4's lack of pointers, memory aliasing, and unstructured control flow (e.g., goto) allow for easier generation of language semantics. Furthermore, using an SMT solver together with translation validation is more precise than randomized testing approaches such as EMI and Csmith because the solver exhaustively searches over all packet inputs to a program to find semantic bugs.

To perform translation validation, we convert P4 programs before and after a compiler pass into logic formulas and assert equivalence of these formulas. To do so, we could have converted P4 programs into C code and then asserted equality using Klee's equivalence-checking mode~\cite{klee}. However, instead, we directly converted P4 programs into logic formulas in Z3~\cite{z3} for two reasons. First, the effort to convert P4 to semantically equivalent C is about the same as producing Z3 formulas directly. The difficulty lies in correctly formalizing all the language constructs of P4, not in the output format. Second, generating Z3 formulas directly gives us more control and allows us to leverage domain-specific techniques to optimize these formulas.

\Para{Model-based testing for black-box compilers.}
Some industry compilers do not have an open specification of their internal program representation or machine code format. In such cases, we cannot use our translation validation technique because it relies on comparing semantics before and after the compiler has transformed the program. Instead, we reuse the semantics we have generated for the input P4 program to determine test cases (i.e., input-output packet pairs) for these random programs. These test cases are then used to directly check the implementations of the P4 programs produced by these compilers. This is effectively model-based testing~\cite{model_based_testing}, with the Z3 semantics serving as a model of the P4 program and the compiler-generated binary being the entity under test.

\subsection{Goals and Non-Goals}
\Para{Find many, but not all bugs.} Our goal is to find many crash and semantic bugs in the P4 compiler, but our tool is not exhaustive. Specifically, we do not intend to build or replace a fully verified compiler like CompCert~\cite{compcert}, given the large labor and time cost associated with such an undertaking with respect to the breadth of P4 back ends. We want to strengthen existing P4 compilers, not write a safe replacement.

\Para{Check the compiler, not the programmer.}
We are not verifying that a particular P4 program is devoid of certain kinds of bugs. This problem is addressed by orthogonal work on P4 program verification~\cite{p4v, bf4, vera, p4-assert, safep4} and P4 testing~\cite{p6}. Although \toolname can in principle be used in for verifying a P4 program, we have not designed it for such use cases. The random programs we generate to find bugs in the P4 compiler are much smaller and more targeted than a typical P4 switch program. Our tool does not need to be able to generate and efficiently solve Z3 formulas for large P4 programs to tease out compiler bugs, although it achieves acceptable performance on large programs (Table~\ref{tab:prog_perf}).

Unlike p4v~\cite{p4v} and Vera~\cite{vera}, whose goal is to provide semantics to find bugs in large programs such as \texttt{switch.p4}, we have developed our semantics for efficient equality checks of diverse, but relatively small, P4 programs. Because of this difference in goals, we believe our semantics cover a broader set of P4 language constructs and corner cases than p4v and Vera---broad enough that we have found bugs in the P4 specification.

\Para{Develop target-independent techniques.} We designed our tools to be as target-independent as possible and specialize them to test the front and mid end of the compiler. While we support restricted forms of back-end testing (\S\ref{sec:symbex}), we do so in a way that allows us to quickly integrate and adapt to new back ends without having to understand detailed target-specific behavior. In particular, we do not cover target-specific semantics such as externs~\cite[\S 4.3]{p416_spec}. We do this by generating programs that are defined in a target-neutral manner with respect to P4\textsubscript{16}'s semantics, i.e., we avoid generating target-specific extern calls.

\Para{Only test mature compilers.} We only test mature compilers such as \pfourc and the corresponding behavioral model\footnote{Both have entered ``permanent beta-status'' since November 2019: \url{https://github.com/p4lang/p4c/issues/2080}} as well as the commercial Tofino compiler. For example, \pfourc supports other back ends such as the eBPF, uBPF, and PSA targets, which are pre-alpha quality and preliminary compiler toolchains. Finding bugs is likely unhelpful for the respective compiler developers at this moment.

\section{Background on P4}
\begin{figure}[t]
    \centering
    \includegraphics[width=0.8\linewidth]{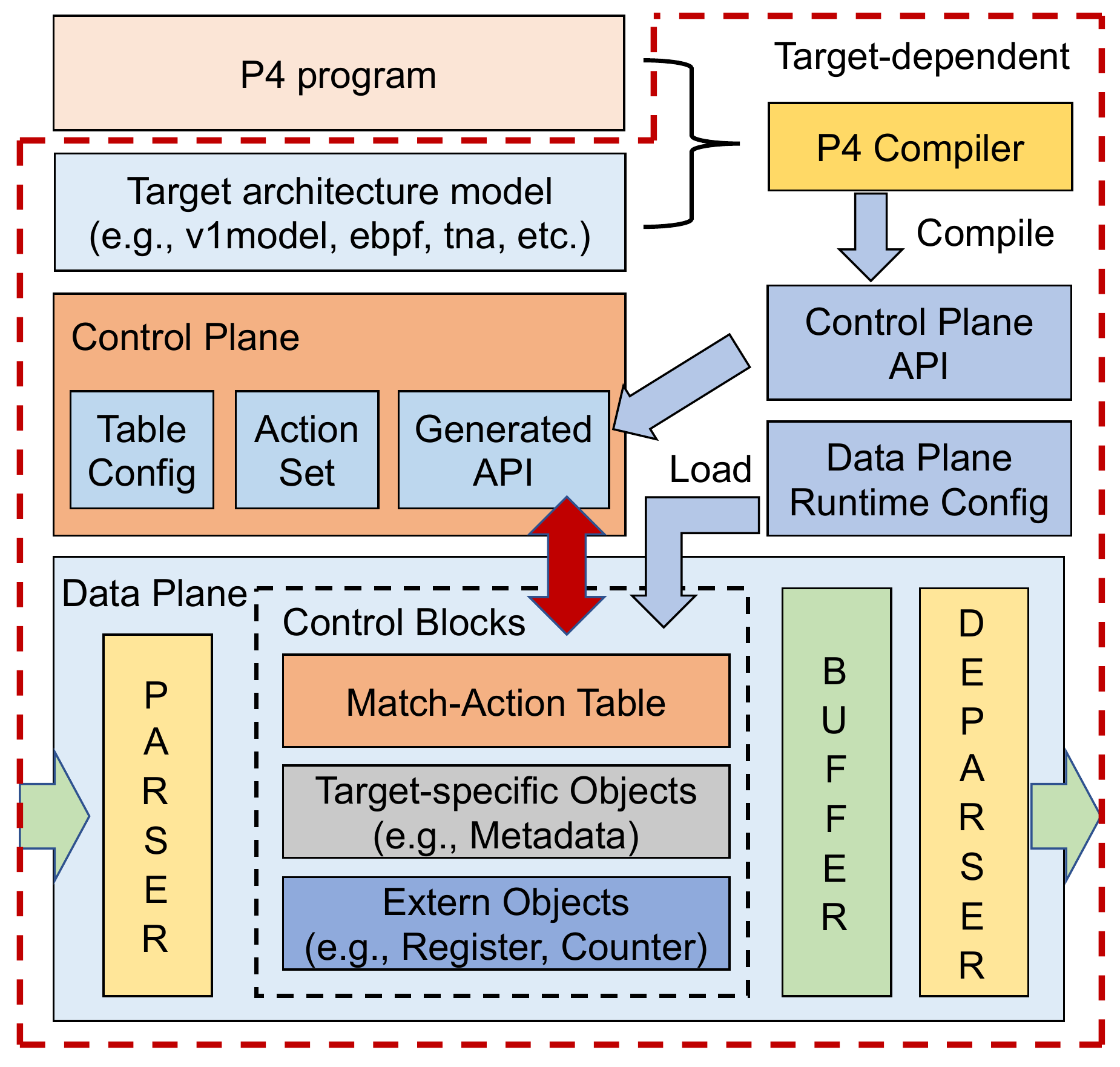}
    \caption{An example P4 compilation model.}
    \label{fig:p4_background}
\end{figure}

P4 is a statically typed DSL designed to describe computations on network packet headers. This paper focuses on P4\textsubscript{16}, the latest version of P4~\cite{p416_spec}. Figure~\ref{fig:p4_background} shows the main P4\textsubscript{16} concepts, explained below.

\Para{Packages and targets.}
A P4 program consists of a set of procedures; each procedure is loaded into a programmable block of the target (e.g., a switch~\cite{tofino} or NIC~\cite{pensando_p4}). These programmable blocks correspond to various subsystems such as the parser or the match-action pipeline. The {\em package} lists the available programmable blocks in a target. One example of a package for a target is the \texttt{v1model}, which models the architecture of a particular \bmv~\cite{bmv2} software switch target, referred to as ``simple switch''~\cite{bmv2_targets}. For simplicity, we will refer to \bmv as the target instead of simple switch.

\Para{P4 compilers.}
A P4\textsubscript{16} compiler translates a P4\textsubscript{16} program and the target package model into target-dependent instructions. These target instructions are combined with the non-programmable blocks (e.g., a fixed scheduler) to form the target's data plane. These instructions also specify how this data plane can be accessed and configured by the control plane (Figure~\ref{fig:p4_background}). \pfourc~\cite{p416} is the official open-source reference compiler infrastructure of the P4\textsubscript{16} language and implements the current state of the specification. \pfourc employs a nanopass design~\cite{nanopass}: a composable library of front- and mid-end compiler passes that perform code analysis, transformation, and optimization on input programs. We analyze these nanopasses using translation validation. 

\Para{Compiler back ends.}
To implement a P4\textsubscript{16} compiler, developers write \pfourc back ends, which use \pfourc's front- and mid-end passes along with their own back-end specific transformations, to translate P4\textsubscript{16} code at the conclusion of the mid end into instructions for their own target. In this paper, we focus on 2 production-grade \pfourc back ends: the Tofino~\cite{tofino} and \bmv~\cite{bmv2} back ends.

\Para{Parsers and control blocks.}
A P4 parser is a finite state machine that transforms an incoming byte sequence received at the target into a structured representation of header definitions. For example, incoming bytes may be parsed as packets containing Ethernet, IP, and TCP/UDP headers. A deparser converts this representation back into a byte sequence. Control blocks describe the per-packet operations that are performed on the input header. These operations are expressed in the form of the core primitives of the language: tables, actions, metadata, and extern objects.

\Para{Tables.}
Tables are objects in the control block similar to a Python dictionary. Table entries are match-action pairs inserted by the network's control plane~\cite{openflow, ethane}. When a table is applied to a packet traversing the control block, its header is compared against the match key of all match-action entries in the table. If any entry's key matches the header, the action associated with the match is executed. Actions are procedures that can modify state and/or input headers.

\Para{Calling conventions.}
P4\textsubscript{16} uses ``copy-in/copy-out''~\cite[\S 6.7]{p416_spec} semantics for method calls. For any callable object in P4, the parameter direction (also known as mode~\cite[\S 8.2]{ada_modes}) explicitly specifies which parameters are read-only and which parameters can be modified, with the modifications persisting after function termination. Modifiable parameters are labelled with the direction \texttt{inout} or \texttt{out} in the definition of the procedure. Read-only values are marked \texttt{in}. At the start of a procedure call, the arguments are copied left-to-right into the associated parameter slots. Parameters with \texttt{out} label remain uninitialized. Once the procedure has terminated, all procedure parameters with the label \texttt{inout} or \texttt{out} are copied back towards the original input arguments. 

\Para{Metadata.}
Metadata is programmer-defined or target-specific data that is associated with a packet header, while it traverses the target. Examples of metadata include the packet input port, packet length, queue depth, or priority; this information is interpreted by the target according to target-specific rules. Metadata can also be modified during the execution of the control block.

\Para{Externs.}
Externs are an extensibility mechanism, which allows targets to describe built-in functionality. Externs are object-like and have methods. Examples include calls to checksum units, hash units, counters, and meters. P4's ``copy-in/copy-out'' semantics allow reasoning about externs to some degree; we can discern which input arguments can take on an arbitrary value and which arguments are read-only.
\section{Random Program Generation}
\label{sec:rand_prog}
Gauntlet's random program generator produces valid P4\textsubscript{16} programs to directly trigger a crash bug. If these programs do not cause a compiler crash they serve as input for our translation validation and model-based testing techniques.

\subsection{Design}
We require diverse input programs to exercise code paths within many compiler passes---and hence bugs in those passes. \pfourc already contains a sample of over 600 programs as part of its test suite. During testing, the reference outputs of each of the test programs are textually compared to the actual outputs after the front- and mid-end passes to check for regressions~\cite[\S 3.4]{p416}. However, this comparison technique is inadequate for semantic bugs. Further, these programs are typically used to test the lexer and parser, not deeper portions of the compiler.

P4Fuzz~\cite{p4fuzz} is a tool that can generate random P4 programs. However, when we tried using P4Fuzz, we found that the programs generated by it are not complex enough to find a large number of new crash or semantic bugs. For example, P4Fuzz generates programs with complex declarations (e.g., \texttt{structs} within \texttt{structs}), but does not generate programs with sufficiently complicated control flow. Hence, it does not cause \pfourc to execute a diverse set of compiler passes. We developed our own generator for random P4 programs that works by generating random abstract syntax trees (ASTs). With this generator we can exercise the majority of language constructs in P4. This leads to diverse test programs covering many combinations of P4 expressions. We can use these test programs to find programs that lead to unexpected crashes.

Gauntlet's random program generator is influenced by Csmith~\cite{csmith} and follows its philosophy of generating only well-formed input programs that pass the lexer, parser, and type checker. The generator grows an AST corresponding to the random program by probabilistically determining what kind of AST node to add to the AST at each step. By adjusting the probabilities of generating each AST node, we can steer the generator towards the language constructs we want to focus on. We can also use these probabilities to keep the size of the average generated program small, in both the number of code lines as well as program paths. With this technique we can find an ample number of semantic bugs while also avoiding programs with too many paths; such ``branch'' programs pose challenges for translation validation and model-based testing.

\Para{Undefined behavior.}
We differ from Csmith in the treatment of undefined behavior. Whereas CSmith tries to avoid generating expressions that lead to undefined behavior, we accommodate such language constructs (e.g., reading from variables that are not initialized). We record the output affected by undefined behavior as part of the logic formulas that we generate from P4 programs during translation validation (\S\ref{sec:undef_behavior}). These formulas allow us to track changes in programs with undefined behavior across compiler passes, which we use to inform compiler developers of suspicious---but not necessarily incorrect---compiler transformations~\cite{stack}.

\subsection{Implementation}
\label{sec:implementation_bludgeon}
We implement our random P4 program generator as extension to \pfourc. The generator uses the intermediate representation (IR) of \pfourc to automatically grow an abstract syntax tree (AST) by expanding branches of the tree at random. For example, a block statement may generate up to (say) 10 statements or declarations, which in turn may result in further sub nodes. The generated IR AST is then converted into a P4 program using \pfourc's \texttt{ToP4} module. Our random program generator can be specialized towards different compiler back ends by providing a skeleton of the back-end-specific P4 package, back-end-specific restrictions, and which package blocks are to be filled in with randomly generated program snippets. We have currently implemented two back ends for our random program generator corresponding to the \bmv~\cite{bmv2_targets} and Tofino~\cite{tofino} targets.

Programs generated by our random program generator are required to be syntactically sound and well-typed. Our aim is not to test if \pfourc can correctly catch syntax and type errors (levels 3 and 4 of Table~\ref{tab:diff_test}). If \pfourc's parser and type checker (correctly) reject a generated program, we consider this to be a bug in our random program generator. For example, if an action parameter has a \texttt{inout} or \texttt{out} qualifier, only writable variables may be passed as arguments.
\section{Translation Validation}
\label{sec:tv}
%%%%%%%%%%%%%%%%%%% FIGURE BEGIN
\begin{figure}[!t]
    \centering
    \includegraphics[width=1\linewidth]{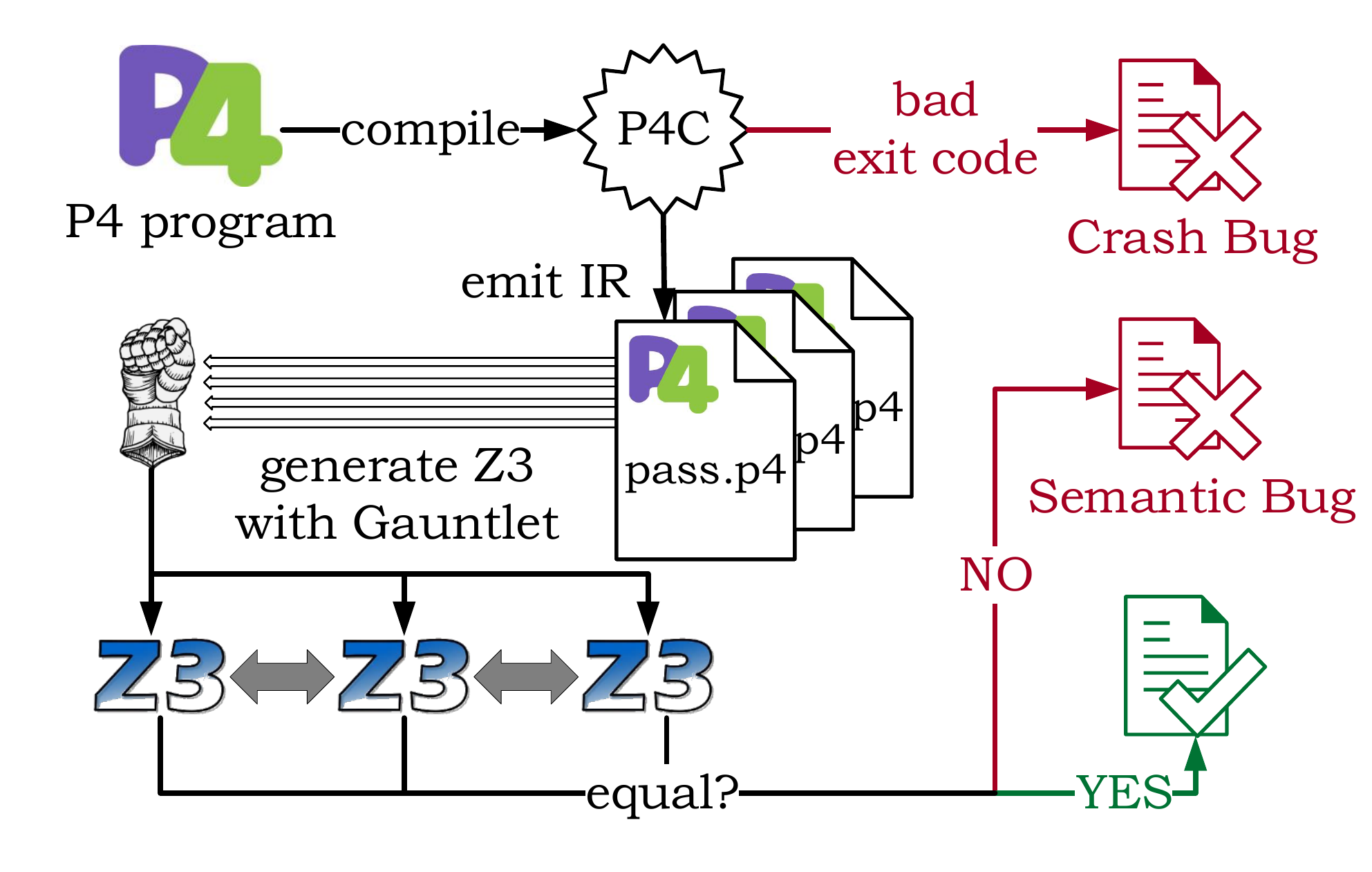}
    \caption{Translation validation in \toolname.\label{fig:workflow_validation}}
\end{figure}
%%%%%%%%%%%%%%%%%%% FIGURE END
To detect semantic bugs, we employ translation validation~\cite{tv_original}, a classic technique from the compiler literature in which an external tool certifies that a particular compiler pass has correctly transformed a given input program. 

\subsection{Design}
To perform translation validation for P4, we developed a symbolic interpreter for the P4\textsubscript{16} language to transform P4 programs into Z3 formulas~\cite{z3}. Figure~\ref{fig:workflow_validation} describes our workflow. To validate a P4 program, the symbolic interpreter converts the program into a Z3 formula capturing its input-output semantics. An equivalence checker then submits the Z3 formulas of a program before and after a compiler pass to the Z3 SMT solver. The solver tries to find an input that violates equivalence of these two formulas. If it finds such an input, this is a semantic bug. Translation validation has two advantages over random testing. First, it can accurately detect subtle differences in program semantics without any knowledge about expected input packets or table entries. Second, when we can access intermediate P4 programs after each compiler pass, we can pinpoint the erroneous pass.

\subsection{Implementation}
\label{sec:implementation_gauntlet}
Like our random program generator, we wrote the interpreter as an extension to \pfourc. We use the IR generated by the \pfourc parser to determine the semantics of a P4 program. Each programmable block of a P4 package represents an independent Z3 formula. For example, the \texttt{v1model} package~\cite{bmv2_targets} of the \bmv back end has 6 different independent programmable blocks: \texttt{Parser}, \texttt{VerifyChecksum}, \texttt{Ingress}, \texttt{Egress}, \texttt{ComputeChecksum}, and \texttt{Deparser}. For each block, we generate a separate Z3 formula.

\Para{Developing the symbolic interpreter.} 
Overall, it took us 5 months of implementation effort until our symbolic interpreter was reliable enough to find new semantic bugs in P4 compilers, instead of encountering false alarms that were actually interpreter bugs. The fact that \pfourc contains a sizeable test suite~\cite[\S 3.4]{p416} was helpful in stress testing our interpreter during development. We started our development process by performing translation validation on programs in the \pfourc test suite. A semantic bug on one of these test programs is probably a false alarm and a bug in our interpreter. This is because it is unlikely that the compiler miscompiles test suite programs. The reference outputs of each test after the front- and mid-end passes are tracked as part of regression testing, and the reference outputs themselves are audited by the compiler developers. We also continuously consulted with the compiler developers to ensure our understanding of the language semantics was correct.

However, we quickly realized that we also needed to generate random programs to achieve coverage and truly stress test our symbolic interpreter. Subsequently, we co-evolved the interpreter with our generator. We attribute part of our success in finding bugs to this development technique, since it forced us to consider many edge cases---more than \pfourc does. The test suite for our interpreter now has over 600 \pfourc tests plus over 100 of our own tests. 

Eventually, our interpreter had become complete and trustworthy enough to perform translation validation for randomly generated programs so as to trigger semantic bugs in \pfourc. After we had detected the first semantic bug, we randomly generated around 10000 programs every week and added the resulting compiler bugs to our backlog. Adding support for new P4 language features as part of random program generation typically first led to a crash in our interpreter. After we fixed our own interpreter, we were frequently able to find new semantic bugs in the P4 compiler that pertained to those language features. Because any of the compiler passes may have bugs, our symbolic interpreter does not rely on any compiler pass of \pfourc. It only relies on the \pfourc parser and the \texttt{ToP4} module to produce P4 code from the IR. Hence, we designed our interpreter to handle any P4 program that successfully passed the \pfourc parser, i.e., before the program is desugared into any normalized form. This allows us to detect semantic bugs in the earliest front-end passes.

%%%%%%%%%%%%%%%%%%% FIGURE BEGIN
\begin{figure}[t]
\begin{subfigure}[b]{\columnwidth}
\begin{lstlisting}[style=P4Style]
struct Hdr { bit<8> a; bit<8> b; }

control ingress(inout Hdr hdr) {
  action assign() { hdr.a = 1; }
  table t {
    key = hdr.a : exact;
    actions = {
      assign();
      NoAction();
    }
    default_action = NoAction();
  }
  apply {
    t.apply();
  }
}
\end{lstlisting}
\caption{Simplified P4 program applying a table.}
\label{fig:p4_to_z3_a}
\end{subfigure}
\hfill
\begin{subfigure}[b]{\columnwidth}
\begin{lstlisting}[style=Z3Style]
Input:  t_table_key, t_action, hdr
Output: hdr_out

hdr_out =
    if (hdr.a == t_table_key) :
        if (1 == t_action) : Hdr(1, hdr.b)
        otherwise : Hdr(hdr.a, hdr.b)
    otherwise : Hdr(hdr.a, hdr.b)
\end{lstlisting}%
\caption{Its semantic interpretation in Z3 shown in functional form.}
\label{fig:p4_to_z3_b}
\end{subfigure}
\caption{A P4 table converted to Z3 semantics.}
\label{fig:p4_to_z3}
\end{figure}

%%%%%%%%%%%%%%%%%%% FIGURE BEGIN

\Para{Converting P4 programs into Z3 formulas.} 
We now describe briefly how we convert a P4 program into a Z3 logic formula. Figure~\ref{fig:p4_to_z3} shows an example. Conceptually, our goal is to represent P4 programs in a functional form so that the input-output behavior of the functional form is identical to the input-output behavior of the P4 program. To determine function inputs and outputs, we use the parameter directions of each P4 package. Parameters with the direction \texttt{inout} and \texttt{out} make up the output Z3 data type of the function whereas  parameters with the \texttt{in} and \texttt{inout} are free Z3 variables that represent the input of the function.

To determine the functional form, the symbolic interpreter traverses each path through the P4 program, maintaining expressions representing path conditions for branching. Once it reaches a portion of the program where execution ends, it stores an if-then-else Z3 expression with the condition set to the path condition and the return value set to a tuple consisting of the \texttt{inout} and \texttt{out} parameters at that point. Ultimately, the interpreter will return a single nested if-then-else Z3 expression, with each branch corresponding to a unique output from the program under a set of conditions. Using this expression we can perform operations such as equivalence checking between two Z3 formulas for translation validation or querying Z3 to provide an output for particular input for test case generation.

\Para{Handling tables.}
The contents of a table are unknown at compile time. Since we want to make sure we cover any possible table content, we interpret match-action pairs in tables symbolically. Figure~\ref{fig:p4_to_z3} describes a simplified example of how \toolname interprets tables within a control block. Per match-action table call, we generate one symbolic match (\texttt{t\_table\_key}) and one symbolic action variable (\texttt{t\_action}), which represent a single match key and its choice of action respectively. We compare the symbolic packet header with the symbolic match key (\texttt{hdr.a == t\_table\_key}). If the expression evaluates to true it implies the execution of a specific action, which is chosen based on the value of the symbolic action index (\texttt{t\_action}). We express this as a series of nested if-then-else statements per action available to the table. Finally, if the key does not match, the default action is selected. For instance, in Figure~\ref{fig:p4_to_z3}, we execute action \texttt{assign} (action id 1) iff the symbolic match variable (\texttt{t\_table\_key}) equals the symbolic header (\texttt{hdr.a}) and the symbolic action variable (\texttt{t\_action}) equals 1. With this encoding we can avoid having to use a separate symbolic match-action pair for every entry in the match-action table, which is a prohibitively large number of symbolic variables.

\Para{Header validity.} The P4\textsubscript{16} specification does not explicitly restrict the behavior of header validity. We model our semantics to align with the implementation in \pfourc. We clarified these assumptions with the compiler and specification maintainers~\cite{p4issue_2323}. If a previously invalid header is marked valid, all fields in that header are initially undefined. If an invalid header is returned in the final output, all fields in the header are set to invalid as well.

\Para{Interpreting function calls.} 
Any \texttt{out} parameter in a function call is initially set undefined. If the function returns, we also generate a new free Z3 variable. In our interpreter, externs are treated as a function call that returns an arbitrary value. In addition, each argument for a parameter that has the label \texttt{inout} and \texttt{out} is set to a new free Z3 variable because the behavior of extern is unknown. Copy-in/copy-out semantics, albeit necessary to control side effects in extern objects, have been a persistent source of bugs in the compiler. A significant portion of the semantic bugs we identified were caused by erroneous passes that perform incorrect argument evaluation and side effect ordering in relation to copy-in/copy-out.

\Para{Checking equivalence between P4 programs.}
\label{sec:implementation_whitebox}
We use \texttt{p4test} to emit a P4 program after each compiler pass. \texttt{p4test} is a \pfourc back end used to test \pfourc. It does not produce any executable output but exercises all the default front- and mid-end passes. We only examine passes that actually modify the input program and ignore any emitted intermediate program that has a hash identical to its predecessor. We explicitly reparse each emitted P4 file to also catch bugs in the parser and the \texttt{ToP4} module.

For an input program A and the transformed output program B after a compiler pass we perform a pair-wise equivalence check for each programmable block. We use our interpreter to retrieve the Z3 formulas for all programmable blocks of the program package and compare each individual block of A to the corresponding block in B. The query for the Z3 solver is a simple inequality. It is satisfiable only if there is a Z3 assignment (e.g., a packet header input or table match-action entry) in which the Z3 formula of A produces a different output from B.

If the inequality query is satisfiable, it produces the assignment that would lead to different results and saves the failed passes for later analysis. With this technique we can precisely pinpoint in which pass a semantic bug may have happened and we can also infer the packet values we need to trigger the bug. If the report turns out to be a false alarm and is not confirmed by compiler developers, this is a bug in our symbolic interpreter, which we fix. The generated Z3 formulas could in principle be very large and checking could take a long time. However, we use quantifier free formulas for the equality check, which can be solved efficiently in Z3~\cite{z3}. Even very large expression trees can be compared under a second.

\Para{Handling undefined behavior.}
\label{sec:undef_behavior}
We track changes in undefined behavior in which the undefined portion of a P4 program has more restricted (less undefined) behavior after a compiler pass. This means we can identify scenarios where the compiler transforms a program fragment based on undefined behavior. While not immediately harmful, such changes might still indicate problematic behavior in the compiler that may be surprising to a programmer~\cite{stack}.

To track undefined behavior, any time a variable is affected by undefined behavior (e.g., a header is set to invalid and then valid) we label that variable ``undefined.'' This undefined variable effectively acts as taint. Every read or write to this undefined variable is tainted. When comparing Z3 formulas before and after a pass, we can choose to replace tainted expressions with concrete values in the formula before a pass.\footnote{We only replace tainted expressions in the ``before'' formula so that we can detect compiler bugs where a previously well-defined expression turns undefined, which is an actual compiler bug, not just an unsafe optimization.} With this, we can determine if a translation validation failure was caused by undefined behavior. If we find a failure based on undefined behavior, we classify it as unstable code~\cite{stack} to avoid confusion with real bugs.
\section{Model-Based Testing}
\label{sec:symbex}
\begin{figure}[t]
    \centering
    \includegraphics[width=1\linewidth]{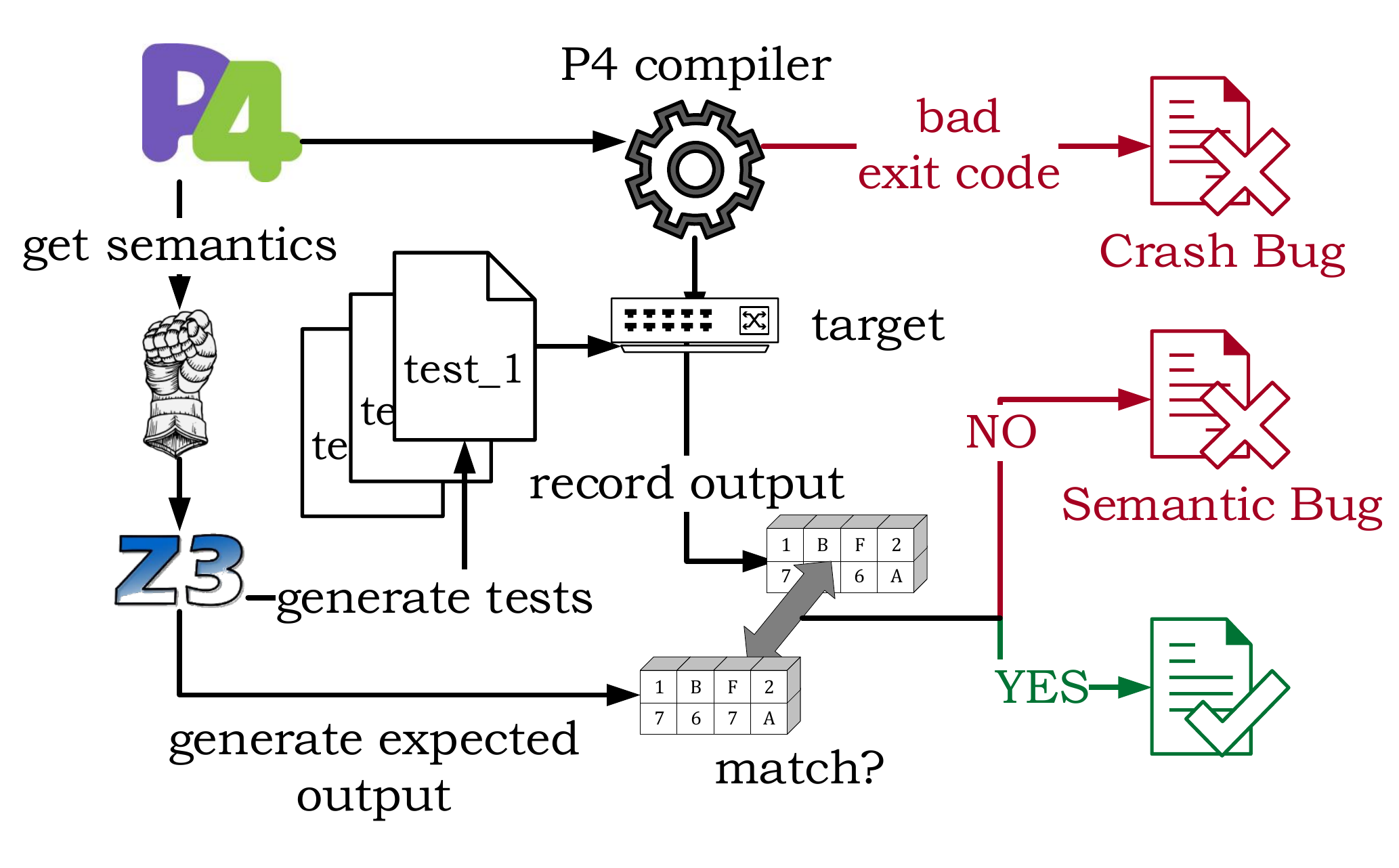}
    \caption{Model-based testing in \toolname.\label{fig:workflow_fuzzing}}
\end{figure}

Our approach to translation validation is applicable only in scenarios where we have access to the P4 IR (and hence the P4 program). This is because it rests on having semantics for P4. This is the case for \pfourc, which has a flag that allows us to emit the input P4 program after every compiler pass as a transformed P4 program~\cite[\S 3.3]{p416}. However, in the back end, a P4 compiler employs back-end-specific passes that translate P4 into proprietary formats. These formats are undocumented, making it hard to provide semantics for them. Hence, to find back-end bugs, we developed a bug-finding approach based on model-based testing~\cite{model_based_testing}.
\subsection{Design}
\label{ssec:semantic_fuzzing}

In this approach, we reuse our symbolic interpreter to produce a Z3 formula of a randomly generated P4 program (Figure~\ref{fig:workflow_fuzzing}). With this Z3 formula, we can produce input packets that traverse unique paths in the P4 program, by generating a path condition for every unique program path and asking Z3 for an input packet that satisfies this path condition. Using the same Z3 formula, we can also determine the output packet for this input packet. Thus, we generate a test case for each path and feed this case into the testing framework of the compiler's target. If the framework reports a mismatch, we know that there is likely a bug. This test technique can identify semantic bugs without requiring access to the P4 program after every intermediate compiler pass. However, unlike the translation validation approach, it is harder to pinpoint the pass causing the bug. This is effectively model-based testing~\cite{model_based_testing} with the Z3 formulas being the model and the compiler output being the system under test.

\subsection{Implementation}
\label{sec:implementation_blackbox}
Model-based testing requires a back-end testing framework that is capable of taking input packets and producing output packets, which can then be matched against the expected output from Z3. We test two back ends: (1) the \bmv back end that uses the simple test framework (STF)~\cite{p4_slides}, which feeds packets to a software test switch and records output packet capture files and (2) the Tofino back end that uses the Packet Test Framework (PTF)~\cite{ptf} to inject and receive packets. We use the Tofino software simulator to check for semantic bugs in Tofino. We initially reconfirmed every semantic bug we found on the Tofino hardware target, but ultimately switched to running only the simulator for faster testing. However, we confirmed all Tofino bugs with the Tofino compiler developers.

\Para{Undefined variables.}
Variables affected by undefined behavior (undefined variables) are difficult to model in model-based-testing because any back end is free to perform arbitrary operations on these variables. We were left with two choices: (1) we could avoid undefined behavior in our P4 programs; (2) alternatively, we could ascribe specific values to undefined variables and check if these values conform with the implementation of the particular target. We picked the second approach because it allows independent testing of compiler optimizations in the face of undefined language constructs.

\Para{Computing input and output for test cases.}
We do not have control over program paths that involve undefined variables because we cannot force a target to assign specific values to such variables. Hence, we add conditions which will cause Z3 to only give us solutions for specific restricted program paths. For any path we can control (e.g., a branch that depends on the value of an input header) we compute all the possible input-output values that lead to a new path through the P4 program. This technique is computationally expensive because the number of paths can be exponential in the length of the program. However, in practice, because our P4 programs have a relatively small number of branches, test-case generation followed by testing on a P4 program still completes quickly. If members of an output header are undefined we mark those bits as ``don't care'' and ignore that portion of the output packet. For any invalid header we omit its member bits from the expected test output.

For every path, we feed path conditions into Z3 and retrieve a candidate set of input-output values that would cause program execution to go down that path. Because there are typically many solutions for these input-output values, we configure the Z3 solver to give us a randomized, non-zero input and its corresponding output value. In some back ends, using zero values by default may mask erroneous behavior. For example, since \bmv initializes any undefined variable with zero, the bug in program~\ref{fig:bug_c} would not have been caught, had we not asked Z3 for a non-zero input-output pair.

\subsection{Limitations}
In contrast to translation validation that runs entirely on a formal logic-based representation of the P4 program, model-based testing has several limitations that are caused by needing to run actual end-to-end tests on real targets.

\Para{Dropped packets in the testing framework.}
A key assumption in the model-based-testing approach is that the generated test cases can actually be fed to the testing framework of the back end. However, the semantics of the generated P4 program do not describe hardware-specific restrictions. For example, some devices impose minimum packet size requirements or drop packets with invalid MAC addresses. More generally, we have found that test cases where the input packets have certain values in their headers can be dropped silently by the back end without generating an output packet. Effectively, there is a mismatch between the Z3 semantics, which says that a certain output packet must be produced and the back end's semantics, which produces no output packet. In such cases, we have had to discard these test cases, reducing the breadth of coverage for testing the compiler.

\Para{Unknown interfaces between programmable blocks.} P4 also does not provide semantics on the treatment of packets in-between the individual control or parser blocks. This is not an issue for translation validation since we compare each programmable block individually. For an end-to-end test, however, we need to know how data is modified between these blocks so that we know what output packet to expect.

\Para{Test case complexity.} Paths with many branches can generate a large number of distinct path conditions. Thus, millions of input-output packet pairs might be generated. Since small programs have sufficed so far for bug finding, we have not run into these issues. In the future, we may need an efficient path selection technique to tease out more complex bugs on closed-source compilers.
\section{Results}
%% TODO: Resume here >>>>>>
\label{sec:results}
We now analyze the P4 compiler bugs found by \toolname. A detailed breakdown can be found at \projecturl. Our main findings are summarized below.
\begin{CompactEnumerate}
    \item We confirmed a total of \bugtotal new, distinct bugs across the \pfourc framework and the \bmv and Tofino P4 compilers. Of these bugs, \bugcrash are crash and \bugsemantic are semantic bugs.
    \item Our efforts led to \specchanges P4 specification changes~\cite[\S A.1]{p416_spec}.
    \item We achieved this in the span of only 8 months of testing with \toolname, and despite only generating random programs from a subset of the P4\textsubscript{16} language.
    \item Model-based testing is effective enough to find semantic bugs in closed-source back ends such as the Tofino compiler, despite us not having access to the internal IR.
    %of the P4 program.
\end{CompactEnumerate}{}

%%%%%%%%%%%%%%%%%%% TABLE BEGIN
 \begin{table}[!t]
 \centering
 %\ra{1.3}
 \begin{small}
 \begin{tabular}{@{}llrrrr@{}}\toprule
 \textbf{Bug Type} & \textbf{Status} & \pfourc & \bmv & Tofino \\ \midrule
  % Crash
 \multirow{3}{*}{Crash}
 & Filed & \bugcrashcompilersubmitted & \bugcrashsimpleswitchsubmitted & \bugcrashtofinosubmitted &   \\ 
 & \textbf{Confirmed} & \bugcrashcompiler & \bugcrashsimpleswitch & \bugcrashtofino &  \\
 & Fixed & \bugcrashcompilerfixed & \bugcrashsimpleswitchfixed & \bugcrashtofinofixed &   \\
 \midrule
 % Semantic
 \multirow{3}{*}{Semantic}
 & Filed & \bugsemanticcompilersubmitted & \bugsemanticsimpleswitchsubmitted & \bugsemantictofinosubmitted &  \\
 & \textbf{Confirmed} & \bugsemanticcompiler & \bugsemanticsimpleswitch & \bugsemantictofino &  \\
 & Fixed & \bugsemanticcompilerfixed & \bugsemanticsimpleswitchfixed & \bugsemantictofinofixed &  \\
 \midrule
 \textbf{Total} & \textbf{\bugtotal} & \textbf{\bugcompiler} & \textbf{\bugsimpleswitch} & \textbf{\bugtofino} \\

 \bottomrule
 \end{tabular}
 \end{small}
 \caption{Bug summary. Unfixed bugs have been assigned.}
 
 \label{table:bug_report}
 \end{table}

 \begin{table}[!t]
 \centering
 %\ra{1.3}
 \begin{small}
 \begin{tabular}{@{}lrrrr@{}}\toprule
 \textbf{Location} & \pfourc & \bmv &  Tofino  & \textbf{Total}\\ \midrule
Front End & \bugfrontend & - & - & \bugfrontend \\ 
Mid End  & \bugmidend & -  & - & \bugmidend \\ 
Back End & - & \bugsimpleswitch & \bugtofino  & \bugbackend\\ 
\midrule
 \textbf{Total} & \bugcompiler & \bugsimpleswitch & \bugtofino & \textbf{\bugtotal} \\

 \bottomrule
 \end{tabular}
 \end{small}
 \caption{Distribution of bugs in the P4 compilers.}
 \label{table:bug_distribution}
 \end{table}
%%%%%%%%%%%%%%%%%%% TABLE END

\subsection{Sources of Bugs}
We distinguish the bugs we found into three primary sources: bugs we found in the common \pfourc framework and bugs we found in the compiler back ends for \bmv and Tofino. Both the \bmv and Tofino back ends use the \pfourc front- and mid-end passes. Hence, most bugs detected in \pfourc also likely apply to these back ends. Note that since the Tofino back end is closed source, we don't know which \pfourc passes it uses.

All semantic bugs in \pfourc were found by translation validation  because we had full access to the compiler IR. Where applicable, we reproduced the semantic bugs using model-based testing and attached the failing input-output packet pair with our bug report. All the semantic bugs in the Tofino compiler were found with model-based testing.

\Para{Distribution of Bugs.}
Table \ref{table:bug_distribution} lists where we identified bugs. The overall majority of bugs were found in the \pfourc front- and mid-end framework, mainly because we concentrated on these areas. The majority of the back end bugs were found in the Tofino compiler. There are two reasons for this. First, the Tofino back end is more complex than \bmv as it compiles for a high-speed hardware target. Second, we did not test the \bmv back end as extensively as other parts of the compiler.

\Para{Bugs in the \pfourc infrastructure.} As Table \ref{table:bug_report} shows, we were able to confirm \bugtotal distinct bugs. \bugcompiler were uncovered in \pfourc, with a comparable distribution of crash bugs (\bugcrashcompiler) and semantic bugs (\bugsemanticcompiler). Initially, the majority of bugs that we found were crash bugs. However, after these crash bugs were fixed, and as our symbolic interpreter became reliable, the semantic bugs began to exceed the crash bugs.

In addition, \specchanges of the bugs we found led to corresponding changes in the specification as they uncovered missing cases or ambiguous behavior because our interpretation of a specific language construct clashed with the interpretation of the compiler developers and language designers. We also continuously checked out the master branch to test the latest compiler improvements for bugs. Many bugs (16 out of \bugcompiler) were caused after recent merges of pull requests during the months in which we used \toolname for testing. \toolname was able to quickly detect these bugs. To catch such bugs as soon they are introduced, the \pfourc developers have now integrated \toolname into \pfourc's continuous integration (CI) pipeline.

\Para{Bugs in the Tofino compiler.} Model-based testing on the Tofino compiler was also successful. We confirmed \bugcrashtofino crash bugs and \bugsemantictofino semantic bugs in the Tofino compiler. These bugs are all distinct from the bugs reported to \pfourc. The majority of bugs present in \pfourc could be reproduced in the Tofino compiler as well, because it uses \pfourc for its front and mid end. Hence, our Tofino bug count does not include any front- and mid-end crash and semantic bugs already present in \pfourc. We also do not include Tofino compiler crashes that were caused by a missed transformation in the \pfourc front end. The Tofino back end was relying on these passes to correctly transform specific P4 expressions. We filed two of these crashes in the Tofino compiler as missed optimization issues in \pfourc. 

\Para{Fixing the bugs.} Out of the \bugtotal new bugs we filed, \bugfixed have been fixed. The remaining bugs have been assigned a developer, but are still open because we filed them very recently, they required a specification change to be resolved first, or they have been de-prioritized in favor of more pressing bug reports. We have received confirmation by the Tofino compiler developers that 8 bugs have already been resolved; the remainder are targeted to be resolved by the next release.

\subsection{Performance on Large P4 Programs}
\begin{table}[!t]
\small
      \centering
      \resizebox{\columnwidth}{!}{
      \begin{tabular}{llll}
      \hline
       Program  & Arch & LoC & Time (mm:ss)\\
      \hline
      \hline
      tna\_simple\_switch.p4 & TNA & 1940 & \textbf{00:05} \\
      \hline
      switch\_tofino\_x0.p4 & TNA & 5751 & \textbf{00:51} \\
      \hline
      switch\_tofino2\_y0.p4 & TNA2 & 6024 & \textbf{00:53} \\
      \hline
      fabric.p4 & V1Model & 958 & \textbf{00:02} \\
      \hline
      switch.p4 (from P4\textsubscript{14}) & V1Model & 5894 & \textbf{10:20} \\
      \hline
      \end{tabular}
      }
      \caption{Time needed to get semantics from a P4\textsubscript{16} program.}
      \label{tab:prog_perf}
\end{table}
We also measured the time \toolname currently requires to generate semantics for several large P4 programs (Table \ref{tab:prog_perf}). Generating semantics is the slowest part of our validation check; comparing the equality of the generated formulas in Z3 is typically fast. We have observed that retrieving semantics for a single pass takes on the order of a minute for a large program. We believe we can substantially improve this performance for two reasons. First, large parts of our semantic interpreter are written in Python as opposed to C++. Second, we currently use a simple state-merging approach for parser branches. This approach does not sufficiently address the scaling challenge of dense branching. When run on \texttt{switch.p4} retrieving semantics takes about 10 minutes. We note, however, that \texttt{switch.p4} is not a representative switch program as the code is autogenerated from old P4\textsubscript{14} code. Programs like \texttt{switch\_tofino\_x0.p4}, which model the data plane of a data center switch, only require a minute per pass.

%%%%%%%%%%%%%%%%%%% TABLE BEGIN

\begin{figure}[!ht]
% % % % % % % % % % % % % % % % % % % % % % % % % % % % % % % % % 
% FIGURE 1
\begin{subfigure}[b]{\columnwidth}
\centering
\begin{lstlisting}[style=P4Style]
control ig(inout Hdr h, ...) {
  apply {
    h.mac_src =
    (h.mac_src > 2 ? 48w1 : 48w2) + h.mac_src;
  }
}
\end{lstlisting}
\caption{A bug caused by a defective pass.}
\label{fig:bug_a}
\end{subfigure}
% % % % % % % % % % % % % % % % % % % % % % % % % % % % % % % % % 
% FIGURE 2
% \hspace*{\fill} % separation between the subfigures
\begin{subfigure}[b]{\columnwidth}
\centering
\begin{lstlisting}[style=P4Style]
control ig(inout Hdr h, ...) {
  apply {
    h.mac_src = (1 << h.modifier) + 8w1;
  }
}
\end{lstlisting}
\caption{A crash in the type checker.}
\label{fig:bug_b}
\end{subfigure}
% % % % % % % % % % % % % % % % % % % % % % % % % % % % % % % % % 
% FIGURE 3
% \hspace*{\fill} % separation between the subfigures
\begin{subfigure}[b]{\columnwidth}
\centering
\begin{lstlisting}[style=P4Style]
control ig(inout Hdr h, ...) {
  apply {
    bool tmp = 1 != 8w2[7:0];
  }
}
\end{lstlisting}
\caption{An incorrect type checking error.}
\label{fig:bug_c}
\end{subfigure}
% % % % % % % % % % % % % % % % % % % % % % % % % % % % % % % % % 
% FIGURE 4
\begin{subfigure}[b]{\columnwidth}
\centering
\begin{lstlisting}[style=P4Style]
control ig(inout Hdr h, ...) {
  action assign_eth_type(inout bit<8> val) {
    h.eth_type[15:8] = 0xFF;
  }
  apply {
    assign_eth_type(h.eth_type[7:0]);
  }
}
\end{lstlisting}
\caption{Incorrect deletion of an assignment.}
\label{fig:bug_d}
\end{subfigure}
% % % % % % % % % % % % % % % % % % % % % % % % % % % % % % % % % 
% FIGURE 5
% \hspace*{\fill} % separation between the subfigures
\begin{subfigure}[b]{\columnwidth}
\centering
\begin{lstlisting}[style=P4Style]
control ig(inout Hdr h, ...) {
  apply {
    h.ipv4.setInvalid();
    h.ipv4.src_addr = 1;
    h.eth.src_addr = h.ipv4.src_addr;
    if (h.eth.src_addr != 1) {
      h.ipv4.setValid();
      h.ipv4.src_addr = 1;
    }
  }
}
\end{lstlisting}
\caption{An unsafe compiler optimization.}
\label{fig:bug_e}
\end{subfigure}
% % % % % % % % % % % % % % % % % % % % % % % % % % % % % % % % % 
% FIGURE 6
% \hspace*{\fill} % separation between the subfigures
\begin{subfigure}[b]{\columnwidth}
\centering
\begin{lstlisting}[style=P4Style]
control ig(inout Hdr h, ...) {
  action assign_and_exit(inout bit<16> val) {
      val = 0xFFFF;
      exit;
  }
  apply {
    assign_and_exit(h.eth_type);
  }
}
\end{lstlisting}
\caption{Incorrect interpretation of exit statements.}
\label{fig:bug_f}
\end{subfigure}

\caption{Examples of bugs that were caught by \toolname.}
\label{fig:bug_samples}
\end{figure}
%%%%%%%%%%%%%%%%%%% TABLE END
\subsection{Deep Dive into Bugs}

\Para{Ripple effects.} A common crash we observed occurs because a compiler pass incorrectly transforms an expression or does not process it at all. Back end compiler developers rely on the front end to correctly transform the IR of the P4 program. But, if a pass misses a language construct it is responsible for, the back end often cannot handle the resulting expression and generates an assertion failure. For example, in program~\ref{fig:bug_a}, the front end \texttt{SideEffectOrdering}~\cite{p4_slides} pass should have converted the conditional operator in line 3 into normal if-then-else control flow. However because of the addition expression, the pass failed to transform the conditional operator, which ultimately caused an assertion to fire in the Tofino back end~\cite{p4issue_2279}. In another case, the \texttt{InlineFunctions}~\cite{p4_slides} pass did not fully inline all functions calls, causing a crash in back ends that were not able to understand function calls and expected them to have been inlined by then~\cite{p4issue_2291}.

\Para{Crashes in the type checker.} Many of the crashes (21 out of \bugcrashcompiler) were in the type checker infrastructure. The code in~\ref{fig:bug_b} shows an expression that crashed type checking~\cite{p4issue_2206}. It is not possible to shift this value since its width is unknown at compile-time. This program was deemed illegal, but the specification did not explicitly forbid it. The type checker tried to infer a type regardless and crashed. This bug also triggered an update to the P4\textsubscript{16} specification~\cite{p4spec_814}. In other cases, the type checker was incorrectly forbidding a valid expression. In example \ref{fig:bug_c}, the program was legal, but because a safety check in the \texttt{StrengthReduction}~\cite{p4_slides} pass was incorrectly implemented, the resulting slice index was overflowing and turned negative, which prompted the type checker to terminate with an error message~\cite{p4issue_2206}.

\Para{Handling side effects.} Side effects from a function operate on the concept of copy-in/copy-out semantics, described earlier. However, these semantics, while seemingly simple, turn out to be hard to implement correctly in the compiler. A particularly tricky case can be seen in \ref{fig:bug_d}~\cite{p4issue_2147}.

In the program, a slice of a variable is passed as an \texttt{inout} parameter. At the same time, a disjoint subset of the variable is assigned within the function. The correct behavior here is to leave the assignment unchanged, and copy back the sliced portion of the variable alone. However, the compiler assumed that the entire variable would be copied back and removed the assignment in line 3, an incorrect optimization.

 A large subset of the semantic bugs we found in \pfourc (at least \bugcopy out of \bugsemanticcompiler) can be traced to incorrect handling of side effects and copy-in/copy-out. Copy-in/copy-out is difficult to handle because for a compiler pass that reorders expressions or statements, side-effects can be translated incorrectly.

\Para{Unstable code.}
Even though the P4\textsubscript{16} language has limited undefined behavior, we also found incidents of unstable code~\cite{stack}. This unstable code conforms with the specification but may lead to instability in specific back end targets. Dumitru et al. also discuss the potential safety consequences of undefined variable access~\cite{exploit_p4}. Program~\ref{fig:bug_e} is a concrete example. The compiler collapses the assignment of line 4 into line 5, setting \texttt{h.eth.src\_addr}, which is still part of a valid header, to 1. All of this is legal behavior, since read and write operations on invalid header values are undefined as part of the P4 specification. The compiler is free to perform these optimizations. However, these changes may cause issues in specific back ends, e.g., back ends in which assignments to invalid headers are no-ops. In this case, the compiler has chosen a particular subset interpretation of undefined behavior, which may clash with the expectations of programmers for that back end. We raised this with the compiler developers, who agreed to print a warning~\cite{p4issue_2323}.

\Para{Consequences of compiler changes.}
Once we started actively monitoring the master branch of \pfourc we observed that many (19 out \bugcompiler) of the bugs we filed in \pfourc were caused by recent merges into master. A notable example is a recent change to the \texttt{Predication}~\cite{p4_slides} pass, which caused at least 6 (1 crash and 5 semantic) new bugs. We caught and filed these bugs quickly during our weekly routine random code generation. The compiler pass has become so complicated that the compiler maintainers are now relying on Gauntlet to ensure correctness ~\cite{pull_2564}. A P4 programmer also filed a bug on this issue~\cite{p4issue_2345}. The report was considered a duplicate because of our earlier reports, highlighting that the bugs we find do affect actual P4 programmers.

\Para{Specification changes.}
Some of our bug reports kicked off larger discussions and changes around the P4 language specification. Our bug reports and questions have led to at least \specchanges distinct specification changes. For example, a concern we had about the validity of uninitialized headers (at what point does a header variable become valid?) led to three clarification pull requests on the specification and a suggestion to propose more fundamental changes for the next language version~\cite{p4spec_849}.

Another prominent example was caused by ambiguity in the specification. In example~\ref{fig:bug_f}, the \texttt{RemoveActionParameters}~\cite{p4_slides} compiler pass moved the statement in line 3 after the exit statement, because the assumption was that exits called within functions ignores the copy-in/copy-out semantics. We instead interpreted exit statements to still respect copy-in/copy-out semantics and caught the discrepancy. This is a significant difference. A packet that traverses the control program could lose all the modifications that have been written to its header, a potential security risk. We filed this as a concern with the open-source community~\cite{p4issue_2225} and our interpretation was deemed reasonable, which required a specification update~\cite{p4spec_823}. The corresponding compiler changes resulted in at least 3 new bugs, which we detected and filed.
%\fruffy{cite?}

\Para{Invalid transformations.} Because \pfourc  provides the option to emit transformed programs after each pass as a valid P4 program, the compiler developers maintain an invariant that each compiler pass in the front and mid end needs to emit syntactically correct P4. We uncovered several bugs with how P4 code is emitted and transformed across compiler passes. We detected these bugs by reparsing each P4 program after it had been emitted by the \texttt{ToP4} compiler module. If the emitted program can not be reparsed, it indicates a bug in one of three compiler components: the \texttt{ToP4} module, the \pfourc parser, or the compiler pass. While these bugs typically do not harm correctness, they affect compiler debugging. Overall, we identified 4 bugs of invalid intermediate P4, all of which were fixed; these 4 are not included in our count of \bugtotal. Additionally, because we reparse P4 after each compiler pass, we found a case where the emitted program being parsed incorrectly was a symptom of a larger bug in the \pfourc parser~\cite{p4issue_2156}.

\subsection{Lessons Learned}
\label{ss:lessons}
\Para{\pfourc debugging support.}
\pfourc has several facilities that were useful for bug finding. The ability to dump the intermediate representation, specify which passes to dump, and the \texttt{ToP4} tool, which converts the P4 IR to P4 programs accelerated our development process. In addition, the compiler has comprehensive assert instrumentation with distinct messages, which we used to identify unique crash bugs and to distinguish them from valid error messages. The AST visitor library in \pfourc allowed us to develop extensions like our random program generator and interpreter.

\pfourc's nanopass architecture, which factors the compiler into a large number of ``thin'' passes, helps with bug fixing, especially for semantic bugs that were narrowed down to one pass by translation validation. A different architecture that has fewer ``thick'' passes would need more developer effort to fix semantic bugs. We also observed that almost all crash bugs were assertion violations where an invariant was violated in a particular compiler pass due to an incorrect or absent compiler transformation from a previous pass. In the absence of such assertions, these crash bugs could have easily manifested as semantic bugs that are harder to detect. 

\Para{Reporting bugs.} This project would not have been possible without the responsiveness and receptiveness of the P4 community. Our questions, concerns, and bug reports were answered within a day and in great detail. The developers were able to even dissect our initial questions and confusions into bug reports, guiding us further in our development effort. We were encouraged to participate in the language design working group that discusses changes to the P4 specification. 

Likewise, when we filed bugs for the closed-source and proprietary Tofino compiler, we found the developers to be receptive and responsive. Still, the pace of bug finding and fixing with the Tofino compiler was slower than the open-source compiler because of two unavoidable reasons. First, we naturally didn't have access to the company bug tracker to assess the life cycle of our bug once it had been filed. Second, the official binary of the Tofino compiler updates less frequently than \pfourc, which can be rebuilt from source after every commit. Hence, we would trigger the same bugs repeatedly in our testing until a new Tofino compiler version with a bug fix was released. Neither of these two problems would manifest, if our tool was to be used internally as part of the compiler development process for Tofino.

\section{Future Work}
\label{sec:future_work}

\Para{New types of bugs.} 
Gauntlet can not find compiler bugs that affect performance or resource usage of generated code. For a switching ASIC that guarantees line-rate performance, the compiler must produce code that consumes a small number of computational and memory units~\cite{chipmunk_sigcomm}. For software targets where line rate performance is not guaranteed, the generated code must have good performance. For example, the P4-eBPF compiler, which converts P4 to eBPF/XDP~\cite{xdp} byte code, occasionally produces code with poor performance~\cite{p4c-xdp}. We are investigating methods that allow us to identify when a compiler pass negatively affects performance and resource usage. We anticipate that handling such bugs would require techniques that are conceptually very different from our methods, which deal with correctness bugs.

\Para{Supporting aggressive compiler optimizations.}
Similar to credible compilation~\cite{credible_compilation}, we plan to repurpose \toolname as an attachable compiler plugin to facilitate development of experimental compiler optimizations. During compilation, if a newly added optimization produces semantically incorrect code, \toolname will notify the compiler to discard the optimization. With this technique, a developer can integrate potentially buggy code into the compiler while still guaranteeing a safe compilation process. However, for the plugin to be useful, \toolname's translation validation needs to be fast enough so that compilation time remains acceptable.

\Para{Extending translation validation to the compiler back end.}
So far we have applied translation validation only to compiler front and mid ends. This is because these passes allow us to dump the P4 program before and after the pass has run, allowing us to compare the before and after programs for equality. The back end is typically proprietary, inaccessible, and uses an opaque intermediate representation. To understand the constraints of these back ends we are currently working with industry compiler developers to integrate translation validation into their compilers. We will develop translation validation techniques that allow us to compare a P4 program’s semantics with the semantics of a back end language that is not P4.

\Para{Long-term study on translation validation in CI.}
Now that translation validation is running as part of the CI pipeline of \pfourc we would like to perform empirical, long-term studies. We want to identify which passes frequently cause semantic issues and understand why they do. We would also like to observe how developer-friendly our tool is. For example to avoid confusing compiler developers, we already had to make sure that \toolname does not report changes in undefined behavior~\cite{pull_2509} or fails gracefully when \toolname does not support a particular language construct~\cite{pull_2451}.

\Para{Automatic test case reduction.} We have not developed an automatic test-case reduction suite (e.g., C-Reduce~\cite{creduce}) and reduce buggy programs in a manual fashion. After our testing pipeline has identified problematic programs in a randomly generated batch, we inspect each P4 program individually. We prune the random P4 program that caused the bug until we get a sufficiently small program that can be attached to a bug report. We are currently automating this process.

\Para{Better coverage of the compiler and  P4\textsubscript{16} language.} While our symbolic interpreter provides semantics for the majority of the P4\textsubscript{16} language constructs, we currently do not generate programs that contain several P4\textsubscript{16} language features: extern calls, method overloading, type definitions, variable bit vectors, run-time indices, match types such as longest prefix or ternary matches, type-inference for generic types in function bodies, annotations, and various custom table properties. We expect that adding most of these will be conceptually straightforward, although adding each language construct is a fair amount of additional engineering. One particular construct that we anticipate being hard to support is externs. While our interpreter includes an extension model to add custom semantics for each extern, extern behavior is very back-end-specific. It is hard to develop accurate semantics for these externs without detailed hardware knowledge of each target. We also do not track how much of the compiler source code we actually cover with our program generator. For future work, we would like to measure the compiler code coverage of a generated P4 program with \texttt{gcov} to understand avenues for improvement.
\section{Related Work}
\label{sec:related_work}
P4K~\cite{p4k} was an effort to formalize the P4 language using the K-framework~\cite{k}. In the process of defining these semantics, the authors found several issues in the P4 specification. P4K supports the use of translation validation similar to our tool. \texttt{netdiff}~\cite{dataplane-equivalence} uses symbolic execution to verify the equivalence of data planes, such as those written in P4. They do so by converting P4 and other data plane programs into the SEFL language~\cite{sefl}, which in turn can be converted to Z3. The Z3 expressions corresponding to different data planes can then be compared for equality. \texttt{netdiff}'s equivalence checking technique is comparable to our translation validation technique. However, neither P4K nor \texttt{netdiff} were explicitly designed for finding compiler bugs. To enable such bug finding, we need both a source of random P4 programs and a translation validation technique to compare intermediate versions of these programs. Further, for some back ends such as the Tofino compiler, translation validation is insufficient, requiring us to use model-based testing instead.

\texttt{p4pktgen}~\cite{p4pktgen} is a P4 test-case generation tool, similar to our model-based testing technique. \texttt{p4pktgen} parses the JSON file generated by the \bmv back end and outputs a Z3 formula, which it uses to create test cases. Using  \texttt{p4pktgen}, the authors were able to find several bugs in how \bmv executes JSON files. However, because it operates on output JSON instead of the input P4 program, unlike \toolname, \texttt{p4pktgen} can not find bugs in intermediate compiler passes.

petr4~\cite{petr4} is a project with the goal of providing independent and complete formal foundations for the P4\textsubscript{16} language. petr4 is complementary to our work. While we are explicitly targeting the official P4\textsubscript{16} compiler and specialized our tools to find bugs during compilation, petr4 aims to find inconsistencies and mistakes in the official P4\textsubscript{16} specification and type system. petr4 provides an interpreter that aims to establish unambiguous semantics for a given P4\textsubscript{16} program. This semantic interpretation can potentially be used to guide the development of our own interpreter semantics.
\section{Conclusion}
\label{sec:conclusion}
This paper presented \toolname, a tool for finding bugs in packet-processing compilers for languages such as P4. \toolname combines random program generation, translation validation, and model-based testing to find both crash and semantic bugs in P4 compilers. It has been highly effective, uncovering \bugtotal new and confirmed bugs. \bugfixed of these have been fixed and the rest have been assigned to a compiler developer. We have open sourced \toolname at \projecturl and it now runs as part of the CI infrastructure of \pfourc.

While we developed \toolname for P4, we believe the core technique that makes \toolname effective is much more general. In particular, \toolname exploits the fact that P4 is a DSL with significant restrictions such as the lack of loops. These restrictions allow us to revive and simplify prior techniques such as translation validation and take them much further in the context of a DSL. For example, to our knowledge, Gauntlet is the first instance of translation validation running as part of a compiler's CI infrastructure. We believe this ability to exploit domain specificity for more effective compiler bug finding will increasingly be applicable to other DSLs beyond P4.
\end{sloppypar}

\section*{Acknowledgements}
\label{sec:ack}
We would like to thank our shepherd, Madan Musuvathi, and the anonymous OSDI reviewers for their valuable feedback. We would also like to thank Amy Ousterhout, Aurojit Panda, Thomas Wies, Michael Walfish, Srinivas Narayana, and Mihai Budiu for their insightful feedback on paper drafts and the project. We are grateful to the P4 compiler team at Barefoot Networks and the open-source P4 community for their feedback and willingness to engage with our bug reports. In particular we would like to thank Mihai Budiu, Nate Foster, Andy Fingerhut, Han Wang, and Antonin Bas for their prompt responses to our many bug reports. We also thank Aatish Varma and Peixuan Gao for experimenting with using AFL for finding bugs in \pfourc as part of their course project.

% deal with buggy flushend
%\atColsBreak{\vskip10pt}
\bibliographystyle{plain}
\bibliography{main}
% \input{aec_appendix}

%%%%%%%%%%%%%%%%%%%%%%%%%%%%%%%%%%%%%%%%%%%%%%%%%%%%%%%%%%%%%%%%%%%%%%%%%%%%%%%%
\end{document}